\documentclass[11pt,a4paper]{article}
\pdfoutput=1
\usepackage[utf8]{inputenc}
\usepackage{jheppub}
\usepackage{amsmath,amsfonts,bm}

\usepackage{verbatim}
\usepackage{graphicx}
\usepackage{color}
\usepackage{bigints}

\newcommand{\ms}{\overline{MS}}

\newcommand{\remove}[1]{}

\begin{document}

\author[a,b]{Joseph Karpie}
\author[a,b]{, Kostas Orginos}
\author[c]{and Savvas Zafeiropoulos}
\affiliation[a]{Department of Physics, The College of William \& Mary, Williamsburg, VA 23187, USA}
\affiliation[b]{Thomas Jefferson National Accelerator Facility, Newport News, VA 23606, USA}
\affiliation[c]{Institute for Theoretical Physics, Heidelberg University, Philosophenweg 12, 69120 Heidelberg, Germany}
\emailAdd{jmkarpie@email.wm.edu }
\emailAdd{kostas@wm.edu}
\emailAdd{ s.zafeiropoulos@thphys.uni-heidelberg.de}

\title{Moments of Ioffe time parton distribution functions from non-local matrix elements}
\abstract{
We examine the relation of moments of parton distribution functions to matrix elements of non-local operators computed in lattice quantum chromodynamics. We argue that after the continuum limit is taken, these non-local matrix elements give access to moments that are finite and can be matched to those defined in the $\ms$ scheme. We demonstrate this fact with a numerical computation of moments through non-local matrix elements in the quenched approximation and we find that these moments are in  agreement with the moments obtained from direct computations of local twist-2 matrix elements in the quenched approximation.}


\maketitle
\flushbottom
\section{Introduction}

It has been  shown by Ji~\cite{Ji:2013dva}  that a numerical computation of parton distribution functions (PDFs) in Euclidean space using lattice QCD may be possible. 
Shortly after the basic idea was introduced several works \cite{Alexandrou:2015rja,Alexandrou:2018pbm,Lin:2014zya,Chen:2017mzz,Broniowski:2017gfp} explored the properties of the methodology as well as introduced alternative approaches~\cite{Ma:2017pxb,Chambers:2017dov,Liu:1993cv,Bali:2018spj}. We refer the reader to~\cite{Lin:2017snn} for a detailed review on the topic.
However, in recent notes Rossi and Testa~\cite{Rossi:2017muf,Rossi:2018zkn} raised a serious question about the validity of such approaches pointing out that the moments obtained from non-local operators are divergent.
As a result  these   approaches may be  impractical  if one does not understand how to subtract non-perturbatively these unwanted effects. Despite the rebuttal in~\cite{Ji:2017rah}, in~\cite{Rossi:2018zkn} it is argued that power divergences present in the moments of Ji's quasi-PDFs are a major obstruction in extracting the full PDF from lattice QCD calculations using this methodology. 

In this work we will discuss potential power divergences in lattice QCD computations of Ioffe time parton distribution functions (PDFs) using the methodology introduced in~\cite{Radyushkin:2017cyf,Orginos:2017kos}.  Ioffe time PDFs are just the Fourier
transforms of the longitudinal momentum fraction $x$  PDFs, where the Ioffe time $\nu$ is the Fourier dual of the momentum fraction $x$.
Ioffe time PDFs are directly related to the matrix elements computed in lattice QCD and therefore analyzing their behavior at finite lattice spacing is simpler. Furthermore, in the approach of~\cite{Radyushkin:2017cyf,Orginos:2017kos}, the light cone limit is taken as $z^2\rightarrow0$ at fixed Ioffe time $\nu$. On the other hand, the moments of PDFs are related to the coefficients of the Taylor expansion of an ultraviolet (UV) finite  matrix element at zero Ioffe time and fixed $z^2$. We use this property to compute the first two moments of PDFs in the $\ms$ scheme and find agreement with direct computations through local matrix elements of twist-2 operators. The issue of power divergent mixing of high dimensional operators with lower dimensional ones (the ``infamous" trace operators) is
a problem (actually an obstruction) if one wants to directly extract from
lattice QCD calculations the PDFs as Fourier Transforms of hadronic matrix
elements of the bi-local operator. On the contrary if, as it is done in the
present paper, lattice data at short-distance are used to fit the operator product expansion (OPE) of the
Ioffe time function in order to extract
moments of PDFs, no question about power divergent mixing arises. A similar issue though in a different context was pointed out in~\cite{Dawson:1997ic}. 
 In Sec.~\ref{sec:defs}, we remind the reader the details of the formalism, we discuss UV divergences and the expansion in moments. In Sec.~\ref{sec:moments} we perform the computation of moments. Sec.~\ref{sec:concs} contains our concluding remarks.

\section{\label{sec:defs}Ioffe time and pseudo parton distribution functions}

Parton distribution functions can be computed from the  hadronic matrix element of a quark bilinear operator with the quark and anti-quark fields separated by a finite distance. 
In the case of  unpolarized,  non-singlet parton densities  the appropriate matrix element is
    \begin{align}
 {\cal M}^\alpha  (z,p) \equiv \langle  p |  \bar \psi (z) \,
 \gamma^\alpha \,  { \hat E} (z,0; A) \tau_3 \psi (0) | p \rangle \,,
\label{Malpha}
\end{align}
where $z$ is an arbitrary separation between the quark fields, $p$ is an arbitrary momentum for the hadron, ${ \hat E}(z,0; A)$ is  the   $z\to 0$ straight Wilson Line 
 in the fundamental representation,  $\tau_3$ is the flavor Pauli matrix,  and $\gamma^\alpha$ is a gamma matrix acting in spin space.
 Lorentz invariance dictates that this matrix element can  be decomposed as
\begin{align} 
{\cal M}^\alpha  (z,p) = &2 p^\alpha  {\cal M}_p (zp, z^2) 
 + z^\alpha  {\cal M}_z (zp,z^2)
\ .
\end{align}

It should be noted that the same matrix element is used to define the collinear PDFs by taking $z$ to be a separation along the light cone. In particular,
with $z=(0, z_{-} , 0, 0)$ in light cone coordinates and \mbox{$\alpha=+$} we obtain
\begin{align} 
{\cal M}^+  (z,p) = &2 p^+  {\cal M}_p (zp, 0) 
\ ,
\end{align}
with the second term dropping out because $z$ does not have a ``$+$" component and noting that $z^2=0$ on the light cone.
Given that  ${\cal M}_p (zp, z^2) $ is Lorentz invariant, we can compute it with any convenient choice of $z$ and $\alpha$, however, if $z$ is not on the light cone,  the
limit $z^2\rightarrow 0$ has to be taken to compute the PDF. This limit is non-trivial as there exists a logarithmic singularity at $z^2=0$. For that reason the PDFs  are defined through factorization of this short distance singularity which results in scale dependent PDFs~\cite{Anikin:1978tj,Radyushkin:2017lvu,Radyushkin:2018cvn}. Furthermore, it is well known that in the limit of $z^2=0$ only the twist-2 contribution survives.
With the above discussion it is easy to see that a non-perturbative computation of the collinear PDFs should start with the computation of the invariant function 
$ {\cal M}_p (zp, z^2) $ from which one can obtain  the twist-2 contribution in the limit  $z^2 \to 0$. Before continuing further, it should be noted that the Lorentz invariant quantity
$zp$ is called Ioffe time $\nu$ in the literature~\cite{Ioffe:1969kf,Braun:1994jq}. 
A convenient choice is   $z=(0, 0, 0, z_3)$ in Cartesian coordinates,  $\alpha$ in the temporal direction i.e. \mbox{$\alpha=0$}, and the hadron momentum $p=(p^0,0,0,p)$. In this case  the  $z^\alpha$-part drops out
 \begin{equation} 
 \langle  p |  \bar \psi (z) \,
 \gamma^0 \,  { \hat E} (z,0; A) \tau_3 \psi (0) | p \rangle \ 
 = 2 p^0 {\cal M}_p (\nu, z_3^2) \,,
 \label{eq:matelem}
\end{equation}
isolating the function ${\cal M}_p (\nu, z^2)$ we seek to compute~\footnote{For simplicity we work with the metric [-, +, +, +] resulting in $z^2= z_3^2$ for $z=(0, 0, 0, z_3)$.}.
Considering a time local matrix element as above also allows for its computation in Euclidean space using lattice QCD. As it is well known~\cite{Briceno:2017cpo}, in this case the matrix element computed in Euclidean space is the same as the Minkowski space counterpart. Therefore, one can analyze  the properties of the above matrix element using the Minkowski metric. Unfortunately, taking $z$ to be space-like, UV singularities arise from the gauge link self energy and end points~\cite{Radyushkin:2017lvu,Radyushkin:2018cvn,Ishikawa:2016znu,Zhang:2018ggy}. It has been shown, that these UV singularities can be factorized in a multiplicative factor and can be renormalized away. A particularly practical proposal for removing these singularities is to consider the ratio
\begin{equation} 
{\mathfrak M}(\nu,z^2)=\frac{ \langle  p |  \bar \psi (z) \,
 \gamma^0 \,  { \hat E} (z,0; A) \tau_3 \psi (0) | p \rangle \ }{\langle  p=0 |  \bar \psi (z) \,
 \gamma^0 \,  { \hat E} (z,0; A) \tau_3 \psi (0) | p=0 \rangle }
 = \frac{ {\cal M}_p (\nu, z^2)}{{\cal M}_p (0, z^2)} \,,
 \label{eq:matelem_red}
\end{equation}
which exactly cancels all these UV singularities leaving behind a finite reduced function for which the regulator can be removed and therefore,  when computed in lattice QCD it 
can be extrapolated to the continuum limit at fixed $\nu$ and $z^2$. This approach was discussed and tested in~\cite{Orginos:2017kos,Karpie:2017bzm}, where it was indeed shown that these UV singularities are absent from the numerical data for ${\mathfrak M}(\nu,z^2)$. Furthermore, with the UV singularities canceled in the ratio the only remaining singularity at  $z^2=0$ is that associated with collinear divergences and therefore the OPE can be used on  ${\mathfrak M}(\nu,z^2)$ resulting in a factorization into a PDF and a perturbatively computable coefficient function.
Therefore we can write
\begin{equation} 
{\mathfrak M}(\nu,z^2)= \int_0^1 d\alpha\, {\cal C}(\alpha,z^2\mu^2,\alpha_s(\mu)) {\cal Q}(\alpha\nu,\mu) \ + \sum_{k=1}^\infty {\cal B}_k(\nu) (z^2)^k\,,
 \label{eq:facrtorization}
\end{equation}
where $\mu$ is the factorization scale in a particular scheme such as $\overline{MS}$, $\alpha_s(\mu^2)$ is the strong coupling constant and  $ {\cal Q}(\nu,\mu)$ is
the Ioffe time PDF in that scheme. Furthermore, there are additional polynomial corrections to the leading order expression that vanish in the limit of $z^2=0$. These corrections are not the same for ${\mathfrak M}(\nu,z^2)$ and ${\cal M}_p (\nu, z^2)$. As it was shown in~\cite{Orginos:2017kos,Karpie:2017bzm}, the polynomial corrections for the case of the reduced function are smaller due to cancellation of certain polynomial terms in the numerator and the denominator. In particular, since the reduced function by construction satisfies  ${\mathfrak M}(0,z^2)=1$, one  can see that $ {\cal B}_k(0) = 0$ and
\begin{equation} 
 \int_0^1 d\alpha\, {\cal C}(\alpha,z^2\mu^2,\alpha_s(\mu)) = 1 \,.
 \label{eq:kernel_norm}
\end{equation}

In the following subsections we proceed with further discussion of UV divergences  and consider the expansion in moments of
the reduced function ${\mathfrak M}(\nu,z^2)$ from where the $\ms$ moments of PDFs can be extracted.

 \subsection{Comment on UV divergences}
 
 The matrix element ${\cal M}_p(\nu,z^2)$ has UV divergences that need to be renormalized before any discussion of the collinear PDFs begins. However, as we said before, these divergences can be renormalized multiplicatively. 
 Given that there exist many ways to treat these divergences one may want to look at the most general case scenario. Let's assume we adopt a particular regularization scheme (a lattice with lattice spacing $a$) and a particular renormalization scheme (e.g. RI-MOM~\cite{Martinelli:1994ty}).
 In this case, the renormalized matrix element at a scale $\mu_{UV}$ (in the case of multiplicative renormalization) would be related to the bare matrix element (denoted by the subscript $a$) by
 \begin{equation}
 \langle  p |  \bar \psi (z) \, \gamma^0 \,  { \hat E} (z,0; A) \tau_3 \psi (0) | p \rangle_a = {\cal Z}(a \mu_{UV})   \langle  p |  \bar \psi (z) \, \gamma^0 \,  { \hat E} (z,0; A) \tau_3 \psi (0) | p \rangle_{\mu_{UV}} \,.
 \end{equation}
 The renormalized matrix element can then be extrapolated to the continuum limit as all UV divergences have been removed.
 It should be noted that the renormalized matrix element depends now on a new scale, the renormalization scale, which takes the place of the lattice spacing. This scale is different from the factorization scale at which the PDFs are defined as one renormalizes UV divergences while the other factorizes collinear divergences present in the PDFs.
 
 In addition, one may define the renormalization group invariant matrix elements (RGI) through the continuum  UV invariant scaling functions $\sigma(\mu_{UV})$ (see for example~\cite{Guagnelli:1999wp}) by
  \begin{equation}
 \langle  p |  \bar \psi (z) \, \gamma^0 \,  { \hat E} (z,0; A) \tau_3 \psi (0) | p \rangle_{RGI}  = \sigma(\mu_{UV}) \langle  p |  \bar \psi (z) \, \gamma^0 \,  { \hat E} (z,0; A) \tau_3 \psi (0) | p \rangle_{\mu_{UV}}\,,
 \end{equation}
where $\sigma(\mu_{UV})$ is regulator independent, but scheme dependent, and the RGI matrix element is scheme and regulator independent. This RGI matrix elements are the objects that need to be analyzed in order to obtain the collinear PDFs. One may obtain them in a variety of methods (e.g. using the RI-MOM scheme~\cite{Alexandrou:2017huk}), and one can discuss their properties without the need for referring to the lattice regulator anymore. Therefore, the ratio
\begin{equation} 
{\mathfrak M}(\nu,z^2)=   \frac{ {\cal M}_p (\nu, z^2)}{{\cal M}_p (0, z^2)} =  \frac{{\cal M}^{RGI}_p (\nu, z^2)}{{\cal M}^{RGI}_p (0, z^2)} \,,
 \label{eq:RGI_ratio}
\end{equation}
is regulator and RG independent and does not have any knowledge of a lattice cut-off that may have been used in order to compute it.
Here  ${\cal M}^{RGI}_p$ is defined by
\begin{equation} 
 \langle  p |  \bar \psi (z) \,
 \gamma^0 \,  { \hat E} (z,0; A) \tau_3 \psi (0) | p \rangle_{RGI} \ 
 = 2 p^0 {\cal M}^{RGI}_p (\nu, z^2) \,.
 \label{eq:matelem_rgi}
\end{equation}
With this discussion it is clear that the expansion of ${\mathfrak M}(\nu,z^2)$ into leading twist parton distribution functions and subleading higher twist contributions
that are suppressed by powers of $z^2$ cannot  fail due to lattice artifacts as it can be performed without the use of a lattice regulator.

In the limit of small space-like $z^2$, the matrix elements of the numerator  and the denominator of the ratio that defines ${\mathfrak M}(\nu,z^2)$ can be expanded using OPE in terms of local non-perturbative, renormalized matrix elements and Wilson coefficients. The Wilson coefficients can in principle be computed  in perturbation theory in a scheme of our choice. The matrix elements can be computed in any regularization and renormalization scheme we desire provided that is matched to the perturbative scheme used to compute the Wilson coeficients.  
If we ignore higher twist effects, then the small $z^2$ expansion of our matrix element would read as
\begin{equation}
{\mathfrak M}(\nu,z^2) = 1 + \frac{1}{2 p^0}\sum_{k=1}^\infty i^k \frac{1}{k!}z_{\alpha_1}\cdots z_{\alpha_k}   c_k(z^2\mu^2)\langle p| {\cal O}_{(k)}^{0\alpha_1 \cdots\alpha_k}| p \rangle_\mu   + {\cal O}(z^2)\,
\label{eq:MSbarOPE}
\end{equation}
where $\langle p| {\cal O}_{(k)}^{0\alpha_1 \cdots\alpha_k}| p \rangle_\mu$ are the familiar spin averaged  twist-2 matrix elements, and 
 $ \mu$  is the factorization scale. The above equation is valid if a factorization scheme without  power divergences is chosen. Such a scheme is $\overline{MS}$.
Note that the factor of $1/(2p^0)$ is inherited from the denominator matrix element that defines ${\mathfrak M}(\nu,z^2)$,
which is also expanded in powers of $z$. However, because the denominator is a zero momentum matrix element, it
does not contain a tower of twist-2 matrix elements which all vanish at zero momentum with the exception of the vector current matrix element which is equal to one in the isovector case that we are considering here. Furthermore, the higher twist effects in the denominator are considered small and are reabsorbed in the ${\cal O}(z^2)$ terms that are omitted. In the previous section we argued that the coefficients
$ {\cal B}_k(\nu)$ in the higher twist expansion are such that $ {\cal B}_k(0) = 0$. It is precisely the reabsorption of the denominator 
polynomial in the $z^2$ term that cancels which brings about $ {\cal B}_k(0) = 0$. Therefore, we can explicitly see how part of the higher twist effects are canceled in the ${\mathfrak M}(\nu,z^2)$.

In conclusion  the ratio ${\mathfrak M}(\nu,z^2)$ is a regularization and renormalization scheme independent quantity that can be expanded using OPE in the
limit of small $z^2$. Part of the polynomial corrections to the leading contribution is canceled by the polynomial terms in $z^2$ arising from the small $z^2$ expansion of the denominator. The above discussion can also be done using a cut-off scheme as a regulator. In this case power divergences will arise in both the computation of the Wilson coefficients and the computation of the matrix elements. 
However,  divergences in the Wilson coefficients  will cancel those of the matrix elements as ${\mathfrak M}(\nu,z^2)$ is regulator independent. This fact was pointed out in an other context in~\cite{Martinelli:1996pk}.

 \section{\label{sec:moments} Computation of Moments}

In this section we discuss the computation of moments of PDFs from  ${\mathfrak M}(\nu,z^2)$ which can be computed on the lattice and
has a well defined continuum limit. Having established that the expansion in moments is well defined in any scheme, we chose to work in the $\ms$ scheme in this section. First one can further simplify the expression in Eq.~(\ref{eq:MSbarOPE}) by replacing the matrix elements with moments  in $\ms$ $a_n(\mu)$. These moments are  defined by
\begin{equation}
\langle p| {\cal O}_{(k)}^{0\alpha_1 \cdots\alpha_k}| p \rangle_\mu = 2 [p^0p^{\alpha_1}\cdots p^{\alpha_k} - \textrm{traces}]_{\textrm{sym}}\, a_{k+1}(\mu)\,,
\end{equation}
where $[\cdots]_{\textrm{sym}}$ stands for symmetrization of indices.
Inserting this  in Eq.~(\ref{eq:MSbarOPE}) we obtain
\begin{equation}
{\mathfrak M}(\nu,z^2) = 1 +  \sum_{k=1}^\infty i^k \frac{1}{k!}\nu^k    c_k(z^2\mu^2)a_{k+1}(\mu)  +  {\cal O}(z^2)\,,
\label{eq:pitdtaylor}
\end{equation}
where the product $p^3 z_3$ has been replaced by the Ioffe time $\nu$. This formula for the moments is derived by the traditional definition 
\begin{equation}
a_n(\mu) =  \int_{-1}^{1} dx\, x^{n-1} \,q(x,\mu)\,,
\end{equation}
where $q(x,\mu)$ is the parton distribution function.
Recalling the definition of Ioffe time PDFs,
\begin{equation}
{\cal Q}(\nu,\mu) = \int_{-1}^{1} dx\, q(x,\mu) e^{ix\nu}\,,
\label{eq:fourier}
\end{equation}
 we can derive  that 
\begin{equation}
(-i)^n\left.\frac{\partial^n\,{\cal Q}(\nu,\mu)}{\partial \nu^n} \right|_{\nu=0}= \int_{-1}^{1} dx\, x^n\,q(x,\mu)  = a_{n+1}(\mu)\,
\label{eq:derQ}
\end{equation}
where $a_{n}(\mu)$, is the $n$-th moment of the parton distribution function.
From this expression and Eq.~(\ref{eq:facrtorization}) we obtain that if one expands in a Taylor series with respect to $\nu$ the reduced function ${\mathfrak M}(\nu,z^2)$,
the coefficients of this Taylor series expansion are the moments of the PDFs up to a multiplicative constant and up to ${\cal O}(z^2)$ higher twist effects.
In other words from Eq.~(\ref{eq:pitdtaylor}) one can introduce ${\bf m}_n$ as,
\begin{equation}
{\bf m}_n\equiv\left.(-i)^n\frac{\partial^n\,{\mathfrak M}(\nu,z^2)}{\partial \nu^n} \right|_{\nu=0}= c_n(z^2\mu^2)
a_{n+1}(\mu) +  {\cal O} (z^2)\,.
\label{eq:kernmom}
\end{equation}
Furthermore, Eq.~(\ref{eq:derQ}) implies that the Wilson coefficients are
\begin{equation}
c_n(z^2\mu^2) =  \int_0^1 d\alpha\, {\cal C}(\alpha,z^2\mu^2,\alpha_s(\mu)) \alpha^n\,.
\label{eq:wilsmom}
 \end{equation}
 Since $ {\cal C}(\alpha,z^2\mu^2,\alpha_s(\mu))$ is known analytically~\cite{Izubuchi:2018srq,Radyushkin:2018cvn,Zhang:2018ggy} to first order in
 $\alpha_s$ in $\ms$, we can easily compute the Wilson coefficients $c_n(z^2\mu^2)$ in $\ms$, by simple integration of Eq.~(\ref{eq:wilsmom}). The leading order $\ms$ expression for $ {\cal C}(\alpha,z^2\mu^2,\alpha_s(\mu)) $ is
 \begin{equation}
 {\cal C}(\alpha,z^2\mu^2,\alpha_s(\mu))  = \delta(1-\alpha) -  \frac{\alpha_s}{2\pi} C_F \left[ B(\alpha)\ln\left(z^2\mu^2\frac{e^{2\gamma_E +1}}{4}\right) + D(\alpha)\right]\,,
 \end{equation}
 where $B(a)$ is the Altarelli-Parisi kernel
\begin{equation}
B(\alpha) = \left[\frac{1+\alpha^2}{1-\alpha}\right]_+\,
\end{equation}
and $D(\alpha)$ is given by
\begin{equation}
D(\alpha) = \left[ 4 \frac{\ln(1-\alpha)}{1-\alpha}  - 2 (1-\alpha) \right]_+\,.
\end{equation}
In the above equations $[ \cdots]_+$ denotes the ``plus prescription''. 
Integrating Eq.~(\ref{eq:wilsmom}) one obtains
\begin{equation}
c_n(z^2\mu^2) =  1  -  \frac{\alpha_s}{2\pi} C_F \left[\gamma_n \ln\left(z^2\mu^2\frac{e^{2\gamma_E +1}}{4}\right) + d_n\right]\,,
 \end{equation}
where
\begin{equation}
\gamma_n = \int_0^1 d\alpha\, B(\alpha) \alpha^n=\frac{3}{2} - \frac{1}{1+n} -  \frac{1}{2+n} - 2 \sum_{k=1}^n \frac{1}{k}\,,
\end{equation}
which  are the well known leading order moments of the Altarelli-Parisi kernel, and 
\begin{equation}
d_n = \int_0^1 d\alpha\, D(\alpha) \alpha^n=2\left[ \left(\sum_{k=1}^n \frac{1}{k}\right)^2 + \frac{2 \pi^2 + n (n + 3) (3 + \pi^2) 
 }{6 (n+1) (n+2)}  -  \psi^{(1)}( n+1)\right]
\,.
\end{equation}
Here $\psi^{(1)}(z)$ is the polygamma function defined as $\psi^{(1)}(z) = d^2 \ln \Gamma(z)/dz^2$ with $\Gamma(z)$ being the $\Gamma$-function.
With the Wilson coefficients computed we can now obtain the $\ms$ moments up to ${\cal O}(\alpha_s^2,z^2)$ directly from the reduced function ${\mathfrak M}(\nu,z^2)$ as
\begin{equation}
a_{n+1}(\mu) = (-i)^n\frac{1}{c_n(z^2\mu^2)}\left.\frac{\partial^n\,{\mathfrak M}(\nu,z^2)}{\partial \nu^n} \right|_{\nu=0}
 +  {\cal O} (z^2,\alpha_s^2)\,.
 \label{eq:msbarMOM}
\end{equation}
Note that in order to do so one needs a precise computation  of ${\mathfrak M}(\nu,z^2)$ in the small $\nu$ region at fixed $z^2$. This is the region in which lattice computations can easily achieve high precision.

\begin{figure}[ht]
\includegraphics[width=0.495\textwidth]{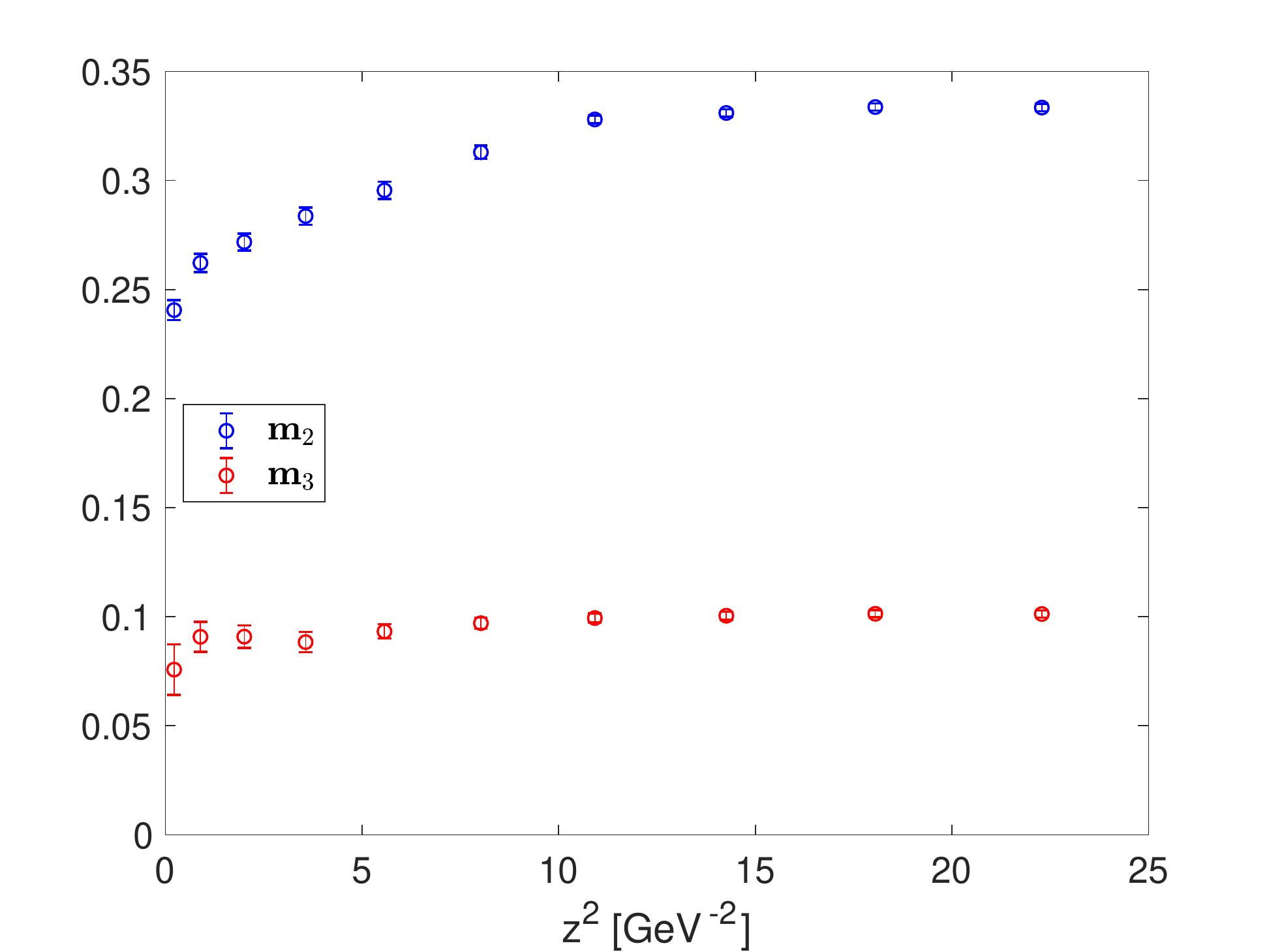}\hspace{.25cm}\includegraphics[width=0.495\textwidth]{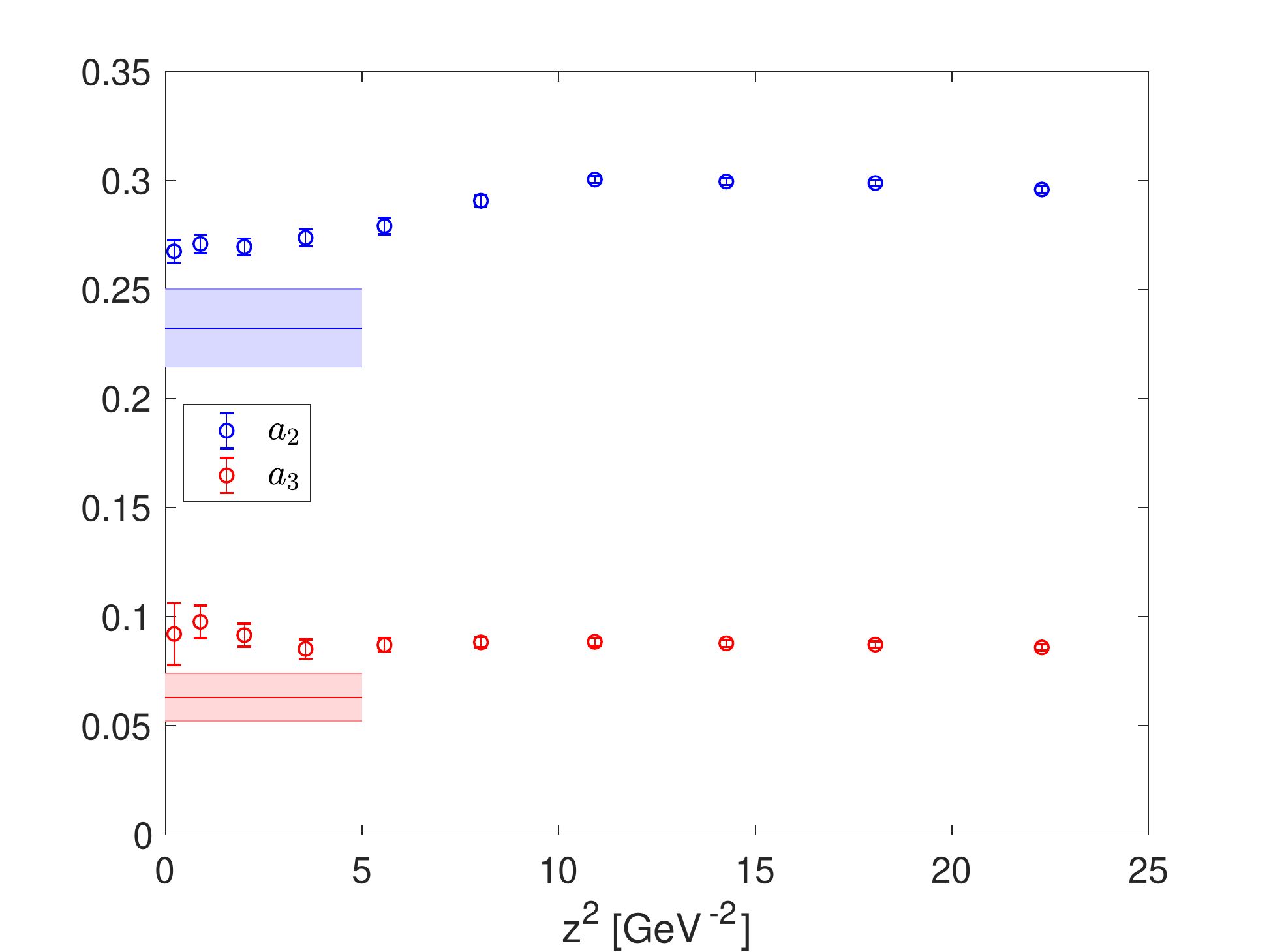}
\caption{{\bf Left:} The first and second derivative of ${\mathfrak M}(\nu,z^2)$ with respect to $\nu$ at $\nu=0$ rescaled by $i$ as defined in Eq.~(\ref{eq:kernmom}). {\bf Right:} The two lowest moments of the isovector unpolarized PDFs at $\mu=3$ GeV versus $z^2$. The shaded error bands  are the QCDSF results for the same pion mass ($\approx 600$ MeV) obtained from~\cite{Gockeler:1995wg} at the same scale $\mu=3$ GeV.   At low $z^2$ the perturbative matching seems to work well as indicated by the independence of the moment on $z^2$. }
\label{fig:msbarMOM}
\end{figure}

 To illustrate this procedure we take as an example a recent quenched QCD 
calculation~\cite{Orginos:2017kos} which can be  compared with the results   from~\cite{Gockeler:1995wg} where the  moments where obtained through direct computations of the matrix elements of twist-2 operators.  In~\cite{Orginos:2017kos}, the reduced isovector Ioffe time pseudo-PDF was computed at a fixed coupling $\beta=6.0$ in quenched QCD using the Wilson gauge action and Clover improved valence fermions. The lattice spacing in this computations is $0.093$ fm. The same quenched theory was also  used in~\cite{Gockeler:1995wg} to study the moments of PDFs from direct computations of the corresponding twist-2 nucleon matrix elements, however in this case Wilson fermions were used for the valence quarks.  The two calculations have very different systematics and most importantly different discretization errors
due to the use of two different valence fermion actions that differ by ${\cal O}(a)$ effects. Nonetheless, it is instructive to check if the moments computed  from the reduced Ioffe time PDF agree within these expected systematic effects with the direct computation. For our comparison the pion mass is set in both cases to $m_\pi\approx 600$ MeV.

On the left panel of Fig.~\ref{fig:msbarMOM}, we plot the left hand side of Eq.~(\ref{eq:kernmom}). These are the derivatives of  ${\mathfrak M}(\nu,z^2)$, rescaled by powers of $i$,  at $\nu=0$.  The  derivatives of  ${\mathfrak M}(\nu,z^2)$ are estimated numerically from  its real and imaginary parts, using finite difference derivatives with ${\cal O}(\nu^4)$ errors, which are of the order of a few percent.  The real part contains only even powers in $\nu$ while the imaginary only odd. This is taken into account in order to simplify the numerical derivative estimator. 

 In the right panel of Fig.~\ref{fig:msbarMOM}, we plot the lowest two moments of the unpolarized isovector  PDF computed using
Eq.~(\ref{eq:msbarMOM}) at  scale $\mu=3\, {\rm GeV}$.  These moments are plotted  as function of $z^2$ used in their extraction. At this scale the perturbative corrections are of order 10\%. At lower scales these corrections become larger as expected and thus the 1-loop matching is expected to break down.  Unfortunately, the available data for ${\mathfrak M}(\nu,z^2)$ do not have sufficient precision to extract higher moments. As we can see at small $z^2$, where the perturbative expansion is expected to work better, the resulting
moment is independent of $z^2$ indicating that the matching formula works well in this region of $z^2$. At higher values of $z^2$,
as expected the perturbative matching breaks down and thus the moment depends on the $z^2$ used in the computation.
Furthermore, the small variation of the derivatives of ${\mathfrak M}(\nu,z^2)$ and the extracted moments  over a wide range of $z^2$
indicates that polynomial corrections are indeed small.
The resulting moments if extrapolated to $z^2=0$, using a constant extrapolation at low $z^2$, are comparable  with the results obtained by QCDSF within the errors of both computations. It should be noted that the QCDSF results were computed at the scale $\mu=\sqrt{2}$ GeV, and for that reason we performed 2-loop running~\cite{Gockeler:2010yr} of their results to the scale $\mu= 3$ GeV where our moments were evaluated.  In addition, the QCDSF computation was performed with unimproved Wilson fermions and therefore has ${\cal O}(a) $ errors. The remaining ${\cal O}(10\%) $ difference between our results and those of QCDSF is expected within the systematic errors associated with perturbative matching, discretization  errors, and the systematics of non-perturbative renomalization performed for the local twist-2 matrix elements. 

Clearly, higher order perturbative matching is needed in order to be able to better estimate the systematic error of the perturbative correction. In addition computations at smaller lattice spacings are required in order to control the lattice spacing errors that should be enhanced for the shortest distance points. Finally, computations at larger volumes will provide a finer grid of $\nu$ values allowing for a more robust extraction of the derivatives of ${\mathfrak M}(\nu,z^2)$. This coupled with better statistics may allow us to extract higher moments as well. Detailed comparison between moments obtained from local and non-local matrix elements would require a control of all the above systematics as well as computations on the same ensembles. This is beyond the scope of the current work and will be done in the future.

\section{\label{sec:concs}Conclusion}
 
 In this work we show that because the ratio ${\mathfrak M}(\nu,z^2)$ introduced   in~\cite{Orginos:2017kos} is free of UV divergences and has a well defined continuum limit,  it can be expanded into moments of parton distribution functions using OPE without any complications arising from power divergences due to the lattice regulator used to compute it non-perturbatively.  In particular, if one expands this matrix element in lattice regularized twist-2 operators,
 which have power divergences due to breaking of the rotational symmetry on the lattice,  the accompanying Wilson coefficients will also have power divergences which exactly cancel the power divergences  of the matrix elements. Therefore, the re-summed OPE expansion
 of the reduced Ioffe time PDF is finite (as expected) due to cancellation of these power divergences order by order in the OPE expansion. 
 Furthermore, using the relation between moments of PDFs and the derivatives of ${\mathfrak M}(\nu,z^2)$ with respect to the Ioffe time $\nu$ at $\nu=0$, it is shown that the moments of PDFs can be computed numerically from ${\mathfrak M}(\nu,z^2)$. The first two moments are found to be in agreement with those computed on the lattice by direct computations of matrix elements in the quenched approximation, within the statistical and systematic errors of the two calculations.
 Given that lattice calculations are much easier in the region of small Ioffe time $\nu$, the methodology we presented here, which focuses on the small $\nu$ region, can lead  to a reliable non-perturbative computation of higher moments of PDFs. At this point we also wish to stress once again an important point. Namely, the issue of power divergent mixing of high dimensional operators with lower dimensional ones is a problem if one wants to directly extract from lattice QCD simulations the PDFs as Fourier transforms of hadronic matrix elements of the bilocal matrix element~\cite{Radyushkin:2018nbf}. However, with the method employed in this study, where lattice data at short-distance are used to fit the OPE of the Ioffe time function in order to extract moments of PDFs, no issues with power divergent mixing arise.

 In the future, our calculations with dynamical quarks will improve on the systematics of the extraction of moments by addressing the sources of systematic errors that arise from the perturbative matching, the lattice spacing effects, as well as the numerical estimation of the derivatives of ${\mathfrak M}(\nu,z^2)$. In addition, computations on larger volumes will allow for better probing of the small Ioffe time $\nu$ region. Such computations are currently  being pursued and will be presented in future publications.
 
\acknowledgments
We thank  Anatoly Radyushkin for especially  stimulating and enlightening discussions  through out this work.
 This work  has been supported
by the U.S. Department of Energy through Grant Number DE- FG02-04ER41302, and 
through contract Number DE-AC05-06OR23177, under which JSA operates the Thomas
Jefferson National Accelerator Facility. SZ acknowledges support by the DFG Collaborative Research Centre SFB 1225 (ISOQUANT).
K.O. acknowledges support in part by STFC consolidated grant ST/P000681/1, and the hospitality from DAMTP and Clare Hall at Cambridge University, where this work was performed. JK acknowledges support from the U.S. Department of Energy, Office of Science, Office of Workforce Development for Teachers and Scientists, Office of Science Graduate Student Research (SCGSR) program. The SCGSR program is administered by the Oak Ridge Institute for Science and Education for the DOE under contract number DE-SC0014664.  This work was performed in part using
computing facilities at the College of William and
Mary which were provided by contributions from the National
Science Foundation (MRI grant PHY-1626177),
the Commonwealth of Virginia Equipment Trust Fund
and the Office of Naval Research. In addition, this
work used resources at NERSC, a DOE Office of Science
User Facility supported by the Office of Science of the
U.S. Department of Energy under Contract \# DE-AC02-
05CH11231.
\begin{center}
\bf{Note added}
\end{center}

While this work was in its completing stages, a preprint by A. Radyushkin~\cite{Radyushkin:2018nbf} appeared on arXiv, where some of the points raised in this manuscript are also discussed. 


\bibliographystyle{apsrev}
\bibliography{ppdfs}

\end{document}